\documentclass[a4paper,11pt]{article}
\usepackage[latin1]{inputenc}
\usepackage[T1]{fontenc}
\usepackage[english]{babel}
\usepackage{graphicx}
\title{Dynamical study of the empty Bianchi type I model in generalised scalar-tensor theory}
\author{Stéphane Fay\\
66 route de la Montée Jaune\\
37510 Savonnières\\
\small{Steph.Fay@Wanadoo.fr}}
\date{1st March 1999}

\begin{document}
\maketitle
\begin{center}
Abstract
\end{center}
A dynamical study of the generalised scalar-tensor theory in the empty Bianchi type I model is made. We use a method from which we derive the sign of the first and second derivatives of the metric functions and examine three different theories that can all tend towards relativistic behaviours at late time. We determine conditions so that the dynamic be in expansion and decelerated at late time.
\vspace{8cm}
\newline
Keys Words: Bianchi I; scalar-tensor theory; dynamical study.
\section{Introduction.}
The scalar-tensor theories of gravitation allow to the gravitational constant to vary. Such a phenomenon happens in a large number of theories which try to unify gravitation with the other interaction forces. In the vacuum case, the most general form of the action of the scalar-tensor theories is written \cite{1} :
\begin{equation} \label{1}
S=\int \left[F(\phi)R-1/2(\nabla \varphi )^{2}-U(\varphi )\right]\sqrt{-g}d^{4}x
\end{equation}
where $\varphi $ is a scalar field, $U(\varphi )$ a potential. We get General Relativity with $F(\varphi )=cte$ and Brans-Dicke theory with $U=0$, $F(\varphi )=\varphi ^{2}/8\omega$ and $\omega=cte$. When $F(\varphi )$ is anatically invertible \cite{13} this action can always be written with a Brans-Dicke scalar field. Putting $\phi=F(\varphi )$ and $\omega(\varphi) =F/\left[2(dF/d\varphi )^{2}\right])$, we get :
\begin{equation} \label{2}
S=\int \left[\phi R-\frac{\omega(\phi)}{\phi}(\nabla\phi)^{2}-U(\phi) \right]\sqrt{-g}d^{4}x
\end{equation}
We will take $U(\phi)=0$ so that we can obtain a Newtonian limit for the weak fields \cite{2}. Techniques to find exact or asymptotic solutions to the field equations derived from action (\ref{2}), with or without matter, in an anisotropic Universe, by means of a conformal transformation, have been described in \cite{2}. Exact solutions and asymptotic behaviours of the scale factor have been analysed for the generalised scalar-tensor theory in FLRW model with matter in \cite{9}. Dynamical studies have been made for Brans-Dicke theory in a FLRW model in \cite{3}\cite{4}\cite{6}. Here, we will work in an empty Bianchi type I Universe. We will introduce new variables, write the field equations with their first derivatives and then perform an  analysis to get analytically the sign of the first and second derivatives of the metric functions, without asymptotic methods, whatever $\omega(\phi)$. Hence we will get the qualitative form of these functions in the Brans-Dicke frame for any time: are they increasing or decreasing, do extrema exist and if so, how many, is there inflation, do they tend towards a power law type, etc. 

In section \ref{s1}, we write the field equations of the vacuum Bianchi type I model with the new variables. In section \ref{s2}, we study particular values of these variables and in section \ref{s3} we describe the method which gives the sign of the first derivatives of the metric functions, depending on the form of $\omega(\phi)$. In section \ref{s4}, we apply our method to three different forms of the coupling function which are all such that $\omega\rightarrow \infty$ and $\omega_{\phi}\omega^{-3}\rightarrow 0$ if we adjust some of their parameters. These two limits ensures that the PPN parameters converge towards values in agreement with the observational data \cite{12}. Thus the different theories, corresponding to different choices of the coupling $\omega(\phi)$, converge towards relativistic behaviours. In section \ref{s5}, we examine the three metric functions and under what conditions they are increasing or decreasing together, etc. In the section (\ref{s6}), we describe the method giving the sign of the second derivatives of the metric functions and examine in which conditions they can be decelerated at late time. We apply our results to the coupling functions of section \ref{s4}.
\section{The field equations} \label{s1}
The metric is:
\begin{equation} \label{2a}
ds^{2}=-dt^{2}+a^{2}(\omega^{1})^{2}+b^{2}(\omega^{2})^{2}+c^{2}(\omega^{3})^{2}
\end{equation}
where the $\omega^{i}$ are the 1-forms of the Bianchi type I model, $t$ the proper time and $a(t)$, $b(t)$, $c(t)$ the metric functions depending on $t$. We define the $\tau$ time as:
\begin{equation} \label{2b}
d\tau=abcdt
\end{equation}
and then, the field equations and the Klein-Gordon equation are written:
\begin{eqnarray}
\frac{a^{,,}}{a}-\frac{a^{,2}}{a^{2}}+\frac{a^{,}}{a}\frac{\phi^{,}}{\phi}-\frac{1}{2}\frac{\omega^{,}}{3+2\omega}\frac{\phi^{,}}{\phi}&=&0 \nonumber \\
\frac{b^{,,}}{b}-\frac{b^{,2}}{b^{2}}+\frac{b^{,}}{b}\frac{\phi^{,}}{\phi}-\frac{1}{2}\frac{\omega^{,}}{3+2\omega}\frac{\phi^{,}}{\phi}&=&0 \label{2c} \\ 
\frac{c^{,,}}{c}-\frac{c^{,2}}{c^{2}}+\frac{c^{,}}{c}\frac{\phi^{,}}{\phi}-\frac{1}{2}\frac{\omega^{,}}{3+2\omega}\frac{\phi^{,}}{\phi}&=&0 \nonumber \\
\frac{a^{,}}{a}\frac{b^{,}}{b}+\frac{a^{,}}{a}\frac{c^{,}}{c}+\frac{b^{,}}{b}\frac{c^{,}}{c}+\frac{\phi^{,}}{\phi}(\frac{a^{,}}{a}+\frac{b^{,}}{b}+\frac{c^{,}}{c})-\frac{\omega}{2}(\frac{\phi^{,}}{\phi})^{2}&=&0 \label{3} \\
\phi^{,,}=-\frac{\omega^{,}\phi^{,}}{3+2\omega} \label{4}\\ \nonumber
\end{eqnarray} \nonumber
We integrate (\ref{4}) and get :
\begin{equation} \label{5}
A\phi^{,}\sqrt{3+2\omega}=1
\end{equation}
$A$ being an integration constant. We see in this last expression that the coupling function must be superior to -3/2 so that the square root is real. We use (\ref{4}) to introduce the second derivative of the scalar field in (\ref{2c}) and put:
\begin{equation} \label{6}
\alpha=\frac{a^{,}}{a}\phi,\mbox{ }\beta=\frac{b^{,}}{b}\phi,\mbox{ }\gamma=\frac{c^{,}}{c}\phi,\mbox{ }\phi^{,}=\Phi
\end{equation}
After integrating, the field equations become:
\begin{eqnarray} \label{7}
\alpha+\frac{1}{2}\Phi=\alpha_{0} \nonumber\\
\beta+\frac{1}{2}\Phi=\beta_{0}  \nonumber\\
\gamma+\frac{1}{2}\Phi=\gamma_{0} \nonumber\\
\alpha\beta+\alpha\gamma+\beta\gamma+\Phi(\alpha+\beta+\gamma)-\frac{1}{4}(A^{-2}-3\Phi^{2})=0
\end{eqnarray} 
$\alpha_{0}$, $\beta_{0}$, $\gamma_{0}$ being integration constants. The constraint imposes the condition :
\begin{equation} \label{8}
\alpha_{0}\beta_{0}+\alpha_{0}\gamma_{0}+\beta_{0}\gamma_{0}=(4A^{2})^{-1}
\end{equation}
The physical solutions are such that the metric functions and the scalar field are positive. Hence, the sign of the variables $\alpha$, $\beta$, $\gamma$ will be the same as the sign of the first derivative of the metric functions. The sign of $\Phi$ will be the same as $\phi^{,}$. Negative scalar fields have already been considered in \cite{5} but it means that, in the Einstein frame, the gravitational constant will be negative. For this reason, many authors deal with positive scalar fields. We will do the same, but the method can easily be extended to negative ones. In what it follows, we will consider only the metric function $a$. What we write for $a$ will be valid for $b$ and $c$. Let us say a few words about exact solutions \cite{2} of the field equations. From (\ref{7}), we can easily show that:
\begin{equation} \label{9}
a=exp(\int \frac{\alpha_{0}}{\phi}d\tau+cte)\phi^{-1/2}
\end{equation}
The scalar field can be calculated by integrating and inverting (\ref{10}):
\begin{equation} \label{10}
d\tau=A\int\sqrt{3+2\omega}d\phi
\end{equation}
Therefore, we can obtain exact solutions of the metric functions for the simple form of the coupling function.

What is the link between the results we will obtain in the $\tau$ time and the behaviours of the metric functions in the $t$ time. Since $a(\tau)=a(\tau(t))=a(t)$, the amplitudes of the metric functions will be the same in both $\tau$ and $t$ times. Moreover as:
\begin{equation}
da/dt=da/d\tau d\tau /dt=da/d\tau (abc)^{-1} \nonumber
\end{equation}
with $abc>0$, the sign of the first derivatives of the metric functions will not be different in $\tau$ or $t$ time. Of course the amplitudes of all the derivatives will be different. While it will always be possible to determine asymptotically the amplitudes of $a^{,}$, this will not be the same for $\dot{a}$.\\
Therefore, as we are mainly interested by the sign of $a^{,}$, $a^{,,}$ and $\ddot{a}$, this is not important. The sign of the second derivatives will be different in both times since
\begin{equation}
d^{2}a/dt^{2}=\ddot{a}=\left[a^{,,}-a^{,}(a^{,}/a+b^{,}/b+c^{,}/c)\right](abc)^{-2} \nonumber
\end{equation}
an overdot denoting differentiation with respect to $t$. 
\\
For these reasons, all that we will say about the sign of the first derivatives will apply to both $t$ and $\tau$ time. Hence, results of section \ref{s2}, \ref{s3}, \ref{s4}, \ref{s5} and in particular table 1 (except the sign of the second derivative of the scalar field which will be different by $\phi^{,,}$ in the $t$ time) will not change in $t$ time since they depends on the sign of constants or first derivative of $\omega$ with respect to $\phi$. In section \ref{s6}, where we will deal with the sign of the second derivatives, we will study separately the sign of $a^{,,}$ and $\ddot{a}$.
\\
Another difference between $\tau$ and $t$ time is that, for instance, $t$ can diverge for a finite value of $\tau$. It can, for instance, transform a Universe that exists during a finite $\tau$ time into a Universe which would exist in an infinite $t$ time. But we will not pay attention to this type of phenomenon in our study. In fact, in most cases, we will use $\phi$ as a time coordinate, particularly in section \ref{s4} and \ref{s6}, and so we will have no need to know the intervals of $\tau$ or $t$.
\section{Study of the first derivative of a metric function.} \label{s2}
We consider the first equation of (\ref{7}). The solution of this equation in the $(\alpha,\Phi)$ plane is represented by a straight line. We have two cases depending on the sign of $\alpha_{0}$, which are represented on graph 1. To describe the variations of the metric function $a$, we have to study the dynamic of a point $(\alpha,\Phi)$ on this straight line so that we know the sign of $\alpha$ and hence, this of $a^{,}$ during the time evolution. The straight line cuts the $\Phi$ axe at $(\alpha,\Phi)=(0,2\alpha_{0})$ and the $\alpha$ axe at $(\alpha,\Phi)=(\alpha_{0},0)$. In $(0,2\alpha_{0})$, we have $\alpha=0$. This means that :

- the metric function $a$ reaches an extrema if the motion of the point $(\alpha,\Phi)$ on the straight line is such that the sign of $\alpha$ change. It is an inflexion point for the metric function, if the motion of the point $(\alpha,\Phi)$ on the straight line changes direction when it reaches $(0,2\alpha_{0})$.

- If the motion of the $(\alpha,\Phi)$ point on the straight line is such that it tend asymptotically towards $(0,2\alpha_{0})$ then a possible explanation is that the scalar field vanishes or that $a\propto\tau$.

In $(\alpha_{0},0)$, the first derivative of the scalar field disappears. We will show below that the scalar field is a monotone function of $\tau$. Hence , $\phi^{,}=0$ can be an inflexion point for $\phi$ in the $\tau$ time if the motion of the point $(\alpha,\Phi)$ changes direction after reaching $(\alpha_{0},0)$. Otherwise it means that the scalar field tends towards a constant. In this last case, we have $\phi\rightarrow \phi^{*}=cte$ and (\ref{5}) shows that $\omega\rightarrow \infty$. If we put $\phi=\phi^{*}$ in the field equations (\ref{7}), the metric functions are written $a=e^{\alpha_{0}\phi^{*-1}(\tau-\tau_{0})}$, $b=e^{\beta_{0}\phi^{*-1}(\tau-\tau_{0})}$, $c=e^{\gamma_{0}\phi^{*-1}(\tau-\tau_{0})}$, and become in the proper time $a=a_{0}t^{p_{1}}$ , $b=b_{0}t^{p_{2}}$ , $c=c_{0}t^{p_{3}}$  with $\sum p_{i}=1$ and $\sum p_{i}^{2}=1-2^{-1}A^{-2}(\alpha_{0}+\beta_{0}+\gamma_{0})^{-2}$. Hence, when $(\alpha,\Phi)\rightarrow (\alpha_{0},0)$, the metric functions tend towards a Kasnerian behaviour.

We can make the following general observations valid in the $\tau$ time: when $\Phi\not\in\left[2\alpha_{0},0\right]$, the more increasing (decreasing)  the scalar field is, the more decreasing (increasing) the metric function will be. When $\Phi\in\left[2\alpha_{0},0\right]$, the scalar field and the metric function increase (decrease) if $\alpha_{0}>0$ ($\alpha_{0}<0$).

The last remark will concern the representation, in the $(\alpha,\Phi)$ plane, of the solutions of the first equation in (7). If we take as a convention that $\sqrt{3+2\omega}>0$, equation (\ref{5}) shows that the sign of $\phi^{,}=\Phi$ depends on the sign of the integration constant $A$. Hence the solution represented in figure 1 by the straight line is physically composed of two separate solutions represented by two half-line, one corresponding to $A>0$ and then $\Phi>0$ and the other to $A<0$ and then $\Phi<0$. So, to the first equation of (\ref{7}) correspond four types of behaviours for the metric function and the scalar field,  depending on the sign of $\alpha_{0}$ and $A$. We will see below that each of them can be split again in two cases depending on the sign of $\Phi^{,}=\phi^{,,}$. These four solutions are illustrated in figure 2. In this figure, \{1\}, \{2\}, \{3\}, \{4\} correspond to the four half-lines which represent the four physically different solutions of the first equation of (\ref{7}). $(\tau_{1})$ and $(\tau_{2})$ represent the finite or infinite values of the time $\tau$ for which $(\alpha,\Phi)$ is equal to $(0,2\alpha_{0})$ and $(\alpha_{0},0)$. In what it follows, we will consider the motion of a point $(\alpha,\Phi)$ on each of the four half-lines. It depends on the form of the coupling function $\omega(\phi)$. To determine it, we need an equation to know how and under which conditions $\Phi$ varies.
\section{Study of the metric functions and scalar field variations depending on the form of $\omega(\phi)$} \label{s3}
We have $d\tau=abcdt$ with $abc>0$. Hence $\tau$ is an increasing function of $t$ and the variations of the metric functions in the $\tau$ time will be the same in the $t$ time. From  (\ref{4}), we deduce the equation which gives the variation of $\Phi$ depending on $\omega(\phi)$ : 
\begin{equation} \label{12}
\Phi^{,}=-\frac{\omega_{\phi}(\phi^{,})^{2}}{3+2\omega}
\end{equation}
with $\omega_{\phi}=\omega^{,}/\phi^{,}=d\omega/d\phi$. $3+2\omega$ is positive since $\omega>-3/2$. Then, the sign of $\Phi^{,}$ depends on the sign of $\omega_{\phi}$ which is independent of the time we consider, namely $\tau$ or $t$ (of course $\Phi^{,}=\phi^{,,}$ and the sign of $\ddot{\phi}$ will be different in the $t$ time. But this is not important here since our final aim is to determine the sign of the first derivatives of the metric functions which does not change in $t$ time). So the results we will find and which depend on the sign of the variations of $\Phi$ will be valid  in both $t$ and $\tau$ times. Hence, if $\omega_{\phi}$ has a constant sign, the motion of the point $(\alpha,\Phi)$ on each half-line will be monotone otherwise its direction will change depending on the sign of $\omega_{\phi}$. We now study the case where $\omega(\phi)$ is a monotone function and get eight different behaviours for the scalar field and the metric function corresponding to the split of each of the 4 previous cases in two cases. First, we consider that the coupling function is an increasing function of the scalar field. Then, $\omega_{\phi}>0$ and from (\ref{12}) we deduce that $\Phi^{,}=\phi^{,,}<0$. Consequently, the motion of the point $(\alpha,\Phi)$ on the half-lines will be such that $\Phi$ decrease. Then, if we are on the half-line \{1\}, the point $(\alpha,\Phi)$ moves from the left to the right. In the same time, $\tau$ increases and then we deduce that $\tau_{1}<\tau_{2}$. On \{1\} we have $\Phi=\phi^{,}>0$ : the scalar field is an increasing function of $\tau$. When $\Phi\rightarrow +\infty$, $\alpha<0$. $\alpha$ remains negative until $(\alpha,\Phi)=(0,2\alpha_{0})$, which means $\tau=\tau_{1}$, and when $\Phi\in\left[0,2\alpha_{0}\right]$, $\alpha$ becomes positive. So, we deduce that the metric function is first decreasing until $\tau=\tau_{1}$ and then increases when $\tau>\tau_{1}$ until $\tau=\tau_{2}$, the value of $\tau$ for which the scalar field becomes a constant: the metric function can have a minimum (but it is not necessarily true as we will see below). The same type of reasoning can be applied when we consider the half-lines \{2\}, \{3\} and \{4\}.

If now we consider that the coupling function is a decreasing function of the scalar field, we have $\omega_{\phi}<0$ and $\Phi^{,}=\phi^{,,}>0$. The point $(\alpha,\Phi)$ moves from the right to the left on each of the four half-lines and we have $\tau_{2}<\tau_{1}$. The same reasoning as in the case $\omega_{\phi}>0$ will hold. Hence we get four more cases. Table 1 summarises these eight cases : we give the sign of the triplet $(\omega_{\phi},A,\alpha_{0})$, independent of the time we consider ($t$ or $\tau$), the scalar field and metric function variations, the direction of the motion of the point on each half-line and we allocate a number for each behaviour.

Another condition has to be fulfilled  in the cases \{1\}, \{1'\}, \{4\}, \{4'\}, to have necessarily an extremum: we have to check if the value $\Phi=2\alpha_{0}$ belongs to the interval in which $\Phi$ varies. For this purpose, we rewrite the equation (\ref{5}) :
\begin{equation} \label{13}
A\Phi\sqrt{3+2\omega}=1
\end{equation}
We determine the interval in which the scalar field $\phi$ varies by imposing the conditions $\sqrt{3+2\omega}>0$ and $\phi>0$. Then from (\ref{13}) we deduce the interval for $\Phi$. The condition for an extremum to exist for the behaviours of type \{1\}, \{1'\}, \{4\} and \{4'\} will be that this last interval contains the value $2\alpha_{0}$. One can also check if the value of the scalar field corresponding to $3+2\omega=(2\alpha_{0}A)^{-2}$ beholds to the interval in which $\phi$ varies.

Now, we consider the case where the coupling function $\omega(\phi)$ is not a monotone function of the scalar field. It means that the sign of $\omega_{\phi}$ will change during the evolution of the dynamic. In the interval of time where $\omega_{\phi}$ will be positive, we will have behaviours of type \{1\}, \{2\}, \{3\} or \{4\} and when it becomes negative the metric function and the scalar field will behave respectively as \{1'\}, \{2'\}, \{3'\} or \{4'\}. Hence, the behaviours of the metric function when the coupling function is not monotone will be a succession of behaviours of type \{i\}+\{i'\}+\{i\}+\{i'\}..., the repetitions of the scheme \{i\}+\{i'\} depending on the number of zero of $\omega_{\phi}$.

Note that to achieve our goal, that is determine the variation (sign of the first derivative) of the metric function, we used quantities such that the second derivative of the scalar field or the amplitude of its first derivative are not invariant when we change time coordinate from $\tau$ to $t$. But these two quantities can always be written as function of $\omega_{\phi}$ or $\omega$ which are independent of time coordinate. Therefore our method is in agreement with the fact that the sign of the first derivative of the metric function is the same in $\tau$ or $t$ time.

In the next section we will consider several forms of the coupling function with a decreasing scalar field, i.e. $A<0$.
\section{Applications.} \label{s4}
We are going to examine the variations of the metric functions with three different forms of the coupling function. The couplings we will consider are interesting for the following reasons. The first coupling is $3+2\omega=\phi_{c}^{2}\phi^{2m}$. When $m>0$ and $\phi\rightarrow \infty$ or $m<0$ and $\phi\rightarrow 0$, $\omega\rightarrow\phi^{2m} \rightarrow\infty$. When $m<-1/4$ and $\phi\rightarrow 0$ or when $m>-1/4$ and $\phi\rightarrow \infty$, $\omega_{\phi}\omega^{-3}\rightarrow 0$. Hence, asymptotically, the theory tends towards relativistic behaviours at late time $(\phi\rightarrow 0)$ when $m<-1/4$. When the scalar field becomes infinite, $\omega(\phi)$ tends towards a power law that corresponds to a power or exponential law for $F(\varphi )$ (see (\ref{1})). Power Law for $\omega(\phi)$ have been studied in \cite{11}. This class of theories is also in agreement with the constraints imposed by the slow logarithmic decrease of the gravitational constant $(dG/dt)G^{-1}$. The two other laws, $2\omega+3=m\mid \ln (\phi/\phi_{0}) \mid ^{-n}$ and $2\omega+3=m\mid1-(\phi/\phi_{0})^{n} \mid^{-1}$ have been studied in \cite{9} in a FLRW Universe.  For the first one, we recover the values of the PPN parameters in General Relativity when $\phi\rightarrow \phi_{0}$ if $n>1/2$, whereas for the second one there is no restriction on the value of the exponent $n$.
\subsection{The theory $3+2\omega=\phi_{c}^{2}\phi^{2m}$} \label{s41}
We have :
\begin{equation} \label{15}
\omega_{\phi}=\phi_{c}^{2}m\phi^{2m-1}
\end{equation}
The expression $3+2\omega$ is positive for all positive values of the scalar field. Hence $\phi$ varies in $\left[0,+\infty\right[$. From (\ref{13}) we deduce that $\Phi$ varies in $\left]-\infty,0\right]$. If $m$ is positive, $\omega_{\phi}>0$ and the metric function behaves as \{2\} and \{4\} whereas if $m$ is negative, $\omega_{\phi}<0$, and it behaves as \{2'\} and \{4'\}. In the Cases \{2\} and \{2'\}, the metric function increases. In the case \{4\} and \{4'\}, from (\ref{13}) we deduce that the metric function has an extremum when the scalar field is equal to $(2\alpha_{0}A\phi_{c})^{1/m}$. This  last  value is  always positive  and then  belongs to  the interval  in which the scalar field varies. We conclude that for the types \{4\} or \{4'\}, the metric function will always have respectively a minimum or a maximum.
\subsection{The theory $2\omega+3=m\mid ln\phi/\phi_{0}\mid^{-n}$.} \label{s42}
We restrict the parameters to $n>0$, $m>0$ so that $2\omega+3$ is positive. We will first consider the case where $\phi>\phi_{0}$ . Then, we can write: 
\begin{equation} \label{16}
2\omega+3=m(ln\phi/\phi_{0})^{-n}
\end{equation}
$\omega_{\phi}$ is always negative and $\Phi\in\left]-\infty,0\right]$. Hence, if $\alpha_{0}>0$, the metric function is increasing. If $\alpha_{0}<0$, the metric function will always have a maximum since $\Phi=2\alpha_{0}$ belongs to the interval where $\Phi$ varies.

If we chose for $\phi$ the interval $\left[0,\phi_{0}\right]$, the metric function has a minimum if $\alpha_{0}<(2A\sqrt{m})^{-1}$. Otherwise, it is increasing.
\subsection{The theory $2\omega+3=m\mid 1-(\phi/\phi_{0})^{n}\mid^{-1}$.} \label{s43}
We restrict the parameters to $n>0$, $m>0$ and will take first $\phi>\phi_{0}$. Hence we have: 
\begin{equation} \label{17}
2\omega+3=m\left[(\phi/\phi_{0})^{n}-1\right]^{-1}
\end{equation}
$\omega_{\phi}$ is always negative. If the integration constant $\alpha_{0}$ is positive, the metric function is increasing, whereas if $\alpha_{0}$ is negative, since $\Phi\in\left]-\infty,0\right]$, the metric function will always have a maximum. If we choose $\phi\in\left[0,\phi_{0}\right]$, the metric function is still increasing when $\alpha_{0}>0$ but have a minimum if $\alpha_{0}<0$.
\section{Behaviour of the three metric functions.} \label{s5}
The graph 3 represents the solutions of the system equations (\ref{7}) on the plane $((\alpha,\beta,\gamma),\Phi)$. We choose without loss of generality $\alpha_{0}<\beta_{0}<\gamma_{0}$. We distinguish four cases : 
\begin{enumerate}
\item If $\Phi>2\gamma_{0}$, all the metric functions are decreasing.
\item If $\Phi\in\left[2\gamma_{0},2\beta_{0}\right]$, the metric function associated with the largest of the integration constants is increasing whereas the two others are still decreasing.
\item If $\Phi\in\left[2\beta_{0},2\alpha_{0}\right]$, the metric function associated with the smallest of the integration constants is the only one to be decreasing.
\item If $\Phi<2\alpha_{0}$, the three metric functions are increasing.
\end{enumerate}
If i constants among $\alpha_{0}$, $\beta_{0}$ and $\gamma_{0}$ are positive, we deduce from figure 3 that when $\phi$ is increasing, whatever the form of $\omega(\phi)$, only the i+1 first cases can exist, when $\phi$ is decreasing, whatever the form of $\omega(\phi)$, only the i+1 last cases can exist. Hence, in the case where $\alpha_{0}$, $\beta_{0}$, $\gamma_{0}$ are positive constant  and $A$ is a negative one, all the metric functions will be increasing whatever the form of $\omega(\phi)$. But, if $\alpha_{0}$, $\beta_{0}$, $\gamma_{0}$  are negative and $A$ positive, all the metric functions will be decreasing. We deduce also that to get three increasing metric functions which tend towards a power law, that is $ ((\alpha,\beta,\gamma),\Phi)\rightarrow ((\alpha_{0},\beta_{0},\gamma_{0}),0)$, when $\tau$(and thus $t$) increases, a necessary condition will be that $\alpha_{0}$, $\beta_{0}$, $\gamma_{0}$  be positive , $A$ and $\omega_{\phi}$ have the same sign.
\section{Study of the second-derivative of the metric function} \label{s6}
In the FLRW models, a positive sign of the first and second derivatives of the scale factor with respect to the cosmic time is the sign of inflation: the expansion in the $t$ time is accelerated. Inflation in generalised scalar-tensor theory and in FLRW models has been studied in \cite{7} and \cite{8}. It seems to be noteworthy that it happens without a cosmological constant or potential. One can talk about inflation only when the second derivatives of the metric functions with respect to $t$ are positives. First, we are going to describe a method giving the sign of the second derivative of the metric function with respect to $\tau$ from the knowledge of $\omega$ and $\omega_{\phi}$. Hence, we will be able to completely determine the qualitative form of the metric function in the $\tau$ time. Second, we apply it and finally we will study the sign of $\ddot{a}$ and obtain conditions to have inflation in Bianchi type I model.
\subsection{Study of $a^{,,}$} \label{s61}
The first spatial component of the field equations is written :
\begin{equation} 
\frac{a^{,,}}{a}=\frac{a^{,2}}{a^{2}}-\frac{a^{,}}{a}\frac{\phi^{,}}{\phi}+\frac{1}{2}\frac{\omega^{,}}{3+2\omega}\frac{\phi^{,}}{\phi} \nonumber
\end{equation}
\begin{equation}
\phi^{2}\frac{a^{,,}}{a}=\alpha^{2}-\alpha\phi^{,}+\frac{1}{2}\frac{\omega_{\phi}}{3+2\omega}\phi^{,2}\phi \nonumber
\end{equation}

But $\phi^{,}=1/(A\sqrt{3+2\omega})$, so we get :
\begin{equation} \label{20}
\phi^{2}\frac{a^{,,}}{a}=\alpha^{2}-\frac{\alpha}{A\sqrt{3+2\omega}}+\frac{1}{2}\frac{\omega_{\phi}}{(3+2\omega)^{2}}\frac{\phi}{A^{2}}
\end{equation}
The sign of the left hand side of (\ref{20}) is the same as $a^{,,}$. The right hand side of equation (\ref{20}) is an equation of degree two in $\alpha$. Hence, we have to know the sign of this equation in order to obtain the sign of $a^{,,}$, i.e. to determine its roots. It is important to recall that $\alpha$ can be expressed as a function of the scalar field. We get : 
\begin{equation} \label{21}
\alpha=\alpha_{0}-\frac{1}{2}\phi^{,}=\alpha_{0}-\frac{1}{2}\frac{1}{A\sqrt{3+2\omega}}
\end{equation}
Now we calculate the determinant of the second degree equation (\ref{20}) : 
\begin{equation} \label{22}
\Delta=\frac{1}{A^{2}(3+2\omega)}-2\frac{\omega_{\phi}}{(3+2\omega)^{2}}\frac{\phi}{A^{2}}
\end{equation}
If $\Delta$ is negative, the second degree equation is positive for all value of $\alpha$ and $a^{,,}$ is positive. Then the dynamic of the metric function is accelerated (this is not inflation since the sign of $a^{,,}$ and $\ddot{a}$ are not necessarily the same). If $\Delta$ is positive, the second degree equation has two real roots $\alpha_{1}$ and $\alpha_{2}$. From (\ref{22}), we deduce that $\Delta<0$ if :
\begin{equation} \label{23}
\omega_{\phi}>\frac{3+2\omega}{2\phi}
\end{equation}
The condition (\ref{23}) will be true for the three metric functions. It does not depend on a specific parameter of one of these functions. Hence, when (\ref{23}) is true, the dynamic of the three metric functions in the $\tau$ time is accelerated. If now we consider $\Delta>0$, we find two roots :
\begin{equation} \label{24}
\alpha_{1,2}=(\frac{1}{A\sqrt{3+2\omega}}\pm\sqrt{\frac{1}{A^{2}(3+2\omega)}-\frac{2\omega_{\phi}}{(3+2\omega)^{2}}\frac{\phi}{A^{2}}})/2
\end{equation}
With the form of the coupling function, one can deduce the conditions so that $a^{,,}$ be positive or negative. By conditions we mean the values of the scalar field and of the different parameters defining the form of the coupling function, which rule the sign of $a^{,,}$. To get this sign, we have to know the sign of:
\begin{equation} \label{25}
\alpha_{1,2}(\phi)-\alpha(\phi)=-\alpha_{0}+(2A\sqrt{3+2\omega})^{-1}\left[2\pm\sqrt{1-2\omega_{\phi}\phi(3+2\omega)^{-1}}\right]
\end{equation}
 When $\alpha_{1}-\alpha$ and $\alpha_{2}-\alpha$ have the same sign, equation (\ref{20}) is positive and thus $a^{,,}$ is positive; otherwise, it means that $\alpha\in\left[\alpha_{2},\alpha_{1}\right]$ and then $a^{,,}$ is negative. At late time, if $\phi_{RG}$ is the value of the scalar field for which $\omega\rightarrow \infty$ and $\omega_{\phi}\omega^{-3}\rightarrow 0$ (which ensures the theory is compatible with the observation) we deduce from (\ref{25}) that a necessary and sufficient condition for the dynamic of the metric function to be decelerated in the $\tau$ time, will be:
\begin{equation} \label{26}
\lim_{\phi\rightarrow \phi_{RG}}\omega_{\phi}<-2\alpha_{0}^{2}A^{2}(3+2\omega)^{2}\phi^{-1}
\end{equation}
\subsection{Applications.} \label{s62}
\subsubsection{Theory $3+2\omega=\phi_{c}^{2}\phi^{2m}$} \label{s621}
Remember that for this form of $3+2\omega$ we have $\phi\in\left[0,+\infty\right[$. We continue to choose $A<0$ in order to have a decreasing scalar field. We get : 
\begin{equation} \label{27}
\alpha=\alpha_{0}-\frac{1}{2}\frac{\phi^{-m}}{A\phi_{c}}
\end{equation}
\begin{equation} \label{28}
\alpha_{1,2}=\frac{\phi^{-m}(1\pm\sqrt{1-2m})}{2A\phi_{c}}
\end{equation}
The condition (\ref{23}) is satisfied when $m>1/2$ : in this case we always have $a^{,,}$, $b^{,,}$ and $c^{,,}$ positive. When $m<1/2$, we have to determine the sign of :
\begin{equation} \label{29}
\alpha_{1,2}-\alpha=\frac{\phi^{-m}(2\pm\sqrt{1-2m})}{2A\phi_{c}}-\alpha_{0}
\end{equation}
We will always have $\alpha_{1}<\alpha_{2}$.

- If $\alpha_{0}=0$, we have $\alpha>\alpha_{1}$ for all values of the scalar field. If $m<-3/2$, from equation (\ref{29}) we deduce that $\alpha_{2}<\alpha<\alpha_{1}$ and thus $\alpha^{,,}<0$. If $m\in\left[-3/2,1/2\right]$, we get $\alpha>\alpha_{1,2}$ and then $a^{,,}>0$. Now we consider general case where $\alpha_{0}\not = 0$.
\begin{itemize}
\item If $m<0$,
	\begin{itemize}
	\item if $\alpha_{0}>0$, when $\phi\rightarrow \infty$, $\alpha>\alpha_{1}$. If $m\in\left[-3/2,0\right]$, 	$\alpha>\alpha_{1,2}$ and if $m<-3/2$, $\alpha\in\left[\alpha_{1},\alpha_{2}\right]$. Then the scalar field 	decreases and when $\phi\rightarrow 0$, $\alpha>\alpha_{1,2}$.
	\item If $\alpha_{0}<0$, when $\phi\rightarrow \infty$, if $m\in\left[-3/2,0\right]$, $\alpha>\alpha_{1,2}$, if 	$m<-3/2$, $\alpha\in\left[\alpha_{1},\alpha_{2}\right]$. When the scalar field decreases and 	$\phi\rightarrow 0$, $\alpha<\alpha_{1,2}$.
	\end{itemize}
	Hence, we deduce that : 
	\begin{itemize}	
	\item If $\alpha_{0}>0$,
		\begin{itemize}			
		\item if $m\in\left[-3/2,0\right]$, we have $a^{,,}>0$,
		\item if $m<-3/2$, we have first $a^{,,}<0$ and then $a^{,,}>0$.
		\end{itemize}
	\item If $\alpha_{0}<0$, 
		\begin{itemize}
		\item if $m\in\left[-3/2,0\right]$, we have $a^{,,}>0$, then $a^{,,}<0$ and finally $a^{,,}>0$,
		\item if $m<-3/2$, we have $a^{,,}<0$ and $a^{,,}>0$.
		\end{itemize}
	\end{itemize}
\item If $m\in\left[0,1/2\right]$,

We will always have $\phi^{-m}(2-1\sqrt{1-2m})>0$. When $\phi\rightarrow \infty$, $\alpha$ is larger than $\alpha_{1,2}$ if $\alpha_{0}>0$ or smaller if $\alpha_{0}<0$. For all value of $\alpha_{0}$, when $\phi$ decreases and tends towards 0, we have $\alpha>\alpha_{1,2}$.

Hence, we deduce that if $\alpha_{0}<0$, first we have $a^{,,}>0$, then $a^{,,}<0$ and at last $a^{,,}>0$. If $\alpha_{0}>0$, we always have $a^{,,}>0$.
\end{itemize}
From the knowledge of $a^{,}$ (see \ref{s42}) and $a^{,,}$ it is now easy to know qualitatively the behaviours of the metric function $a$, depending on its different parameters $\alpha_{0}$ and $m$. We deduce from our qualitative analysis that:
\begin{itemize}
\item When \underline{$m\in\left[0,1/2\right]$} and $\alpha_{0}>0$, the metric function is increasing and accelerated. When $\alpha_{0}<0$, the metric function has a minimum. The branch before the minimum is accelerated whereas the branch after the minimum has an inflexion point and is accelerated in late time. 
\item When \underline{$m>1/2$}, the dynamic of the metric function is always accelerated.
\item When \underline{$m<0$} and $\alpha_{0}>0$, the metric function increases. It is accelerated if $m\in\left[-3/2,0\right]$. If $m<-3/2$, it is first decelerated and then accelerated: the metric function has an inflexion point. If $\alpha_{0}<0$, the metric function has a maximum. If $m\in\left[-3/2,0\right]$, the dynamic is accelerated in both late and early times whereas if $m<-3/2$, it is decelerated in early time and accelerated in late time.
\end{itemize}
Note that one can always obtain the value of the scalar field for which the sign of $a^{,,}$ changes by writing $\alpha_{1,2}-\alpha=0$. We see that the theory $3+2\omega=\phi_{c}^{2}\phi^{2m}$ is always accelerated in late time in accordance with the relation (\ref{26}).
\subsubsection{The theory $2\omega+3=m\mid \ln \phi/\phi_{0}\mid^{-n}$.}  \label{s622}
Here, we consider only the interval $\left[\phi_{0},\infty\right[$ for the scalar field, $\omega_{\phi}$ is always negative and then $\Delta$ is always positive. We have:
\begin{equation} \label{30}
\alpha_{1,2}-\alpha=-\alpha_{0}+(2A\sqrt{m})^{-1}(\ln \phi/\phi_{0})^{n/2}(2\pm\sqrt{1+n\phi_{0}\ln (\phi/\phi_{0})^{-1}})
\end{equation}
When \underline{$\alpha_{0}>0$}, in early time, $\phi\rightarrow \infty$ and $\alpha>\alpha_{1,2}$. Then, at late time, when $\phi\rightarrow \phi_{0}$, if $n>1$, we have again $\alpha>\alpha_{1,2}$ and then the metric function increases and is accelerated whereas if $n\in\left[0,1\right]$, we have $\alpha\in\left[\alpha_{1},\alpha_{2}\right]$. Then, the metric function increases but have an inflexion point. It is decelerated at late time.
\\
When \underline{$\alpha_{0}<0$}, the metric function has a maximum. If $n>1$, the dynamic is both accelerated in early and late time whereas if $n\in\left[0,1\right]$, it is just accelerated in early time.
\subsubsection{The theory $2\omega+3=m\mid 1-(\phi/\phi_{0})^{n}\mid^{-1}$.}  \label{s623}
Here again we consider the same interval for $\phi$ and $\Delta$ will be always positive. We have:
\begin{equation} \label{31}
\alpha_{1,2}-\alpha=-\alpha_{0}+(2A\sqrt{m})^{-1}\sqrt{(\phi/\phi_{0})^{-n}-1}(2\pm\sqrt{1+n(\phi/\phi_{0})^{n}/\left[(\phi/\phi_{0})^{n}-1\right]})
\end{equation}
We get two important values for $n$: $n=3$ or $n=4A^{2}\alpha_{0}^{2}m$.
\begin{itemize}
\item When $\alpha_{0}>0$, the metric function is increasing and its behaviour is accelerated if $n<(3,4A^{2}\alpha_{0}^{2}m)$ or decelerated if $n>(3,4A^{2}\alpha_{0}^{2}m)$. If  the value of $n$ is between $n=3$ and $n=4A^{2}\alpha_{0}^{2}m$, the metric function has an inflexion point and the dynamic will be accelerated at late time if $3<4A^{2}\alpha_{0}^{2}m$ or decelerated if $3>4A^{2}\alpha_{0}^{2}m$.
\item When $\alpha_{0}<0$, the metric function has a maximum. Its behaviour is decelerated if $n>(3,4A^{2}\alpha_{0}^{2}m)$. If $n<(3,4A^{2}\alpha_{0}^{2}m)$ , the dynamic is accelerated at both late and early times. If  the value of $n$ is between $n=3$ and $n=4A^{2}\alpha_{0}^{2}m$, the dynamic is decelerated at early time when $3<4A^{2}\alpha_{0}^{2}m$ and becomes accelerated whereas when $3>4A^{2}\alpha_{0}^{2}m$, it is first accelerated and then decelerated at late time.
\end{itemize}
In all the applications one can prove that the behaviours of $a^{,,}$ at early and late times are continuous. The sign of $a^{,,}$ does not change between the late and early times because $(\alpha_{1,2}-\alpha)^{,}$ vanish for only one value of $\phi$ in the intervals in which the parameters of the three theories and the scalar field are allowed to vary. If it was not the case, the sign of this last expression would vanish for, at least, two values of the scalar field.

In the next subsection we will talk about the second derivative of the metric function in $t$ time. For the sake of simplicity (the sign of the second derivative can change more than twice in $t$ time) we will not study the behaviour of these theories in the $t$ time (qualitatively, only the sign of the second derivative changes). Moreover, to do this we must carry out numerical computations as we will see, that seems diverge from our goal, i.e. make a general study of the dynamic whatever the coupling function.
\subsection{Study of $\ddot{a}$.} \label{s63}
Here, when $\ddot{a}$ and the first derivative are positives one can speak about inflation. We have:
\begin{equation} \label{31a}
\frac{\ddot{a}}{a}=\left[\frac{a^{,,}}{a}-\frac{a^{,2}}{a^{2}}-\frac{a^{,}}{a}(\frac{b^{,}}{b}+\frac{c^{,}}{c})\right](abc)^{-2} \nonumber
\end{equation}
The relations (\ref{5}) and (\ref{20}) imply:
\begin{equation} \label{32}
\frac{\ddot{a}}{a}(abc)^{2}\phi^{2}=\frac{1}{2}\frac{\omega_{\phi}}{(3+2\omega)^{2}}\frac{\phi}{A^{2}}-\alpha(\beta_{0}+\gamma_{0})
\end{equation}
This is an equation of first degree for $\alpha$. Its solution is:
\begin{equation}
\alpha_{3}=\frac{1}{2}\frac{\omega_{\phi}}{(3+2\omega)^{2}}\frac{\phi}{A^{2}}(\beta_{0}+\gamma_{0})^{-1} \nonumber
\end{equation}
We use equation (\ref{21}) to write:
\begin{equation} \label{33}
\alpha-\alpha_{3}=\alpha_{0}-\frac{1}{2}\frac{1}{A\sqrt{3+2\omega}}-\frac{1}{2}\frac{\omega_{\phi}}{(3+2\omega)^{2}}\frac{\phi}{A^{2}}(\beta_{0}+\gamma_{0})^{-1}
\end{equation}
Then, one has to solve $\alpha-\alpha_{3}=0$  for $\phi$ so that we can determine the sign of this last expression for different intervals of the scalar field. This is not an easy task and to study the theories of the last subsection, we would need numerical computation. In a general manner, to simplify the resolution, one can notice that equation (\ref{33}) is a third degree equation for $(3+2\omega)^{-1/2}$.
Then, $\ddot{a}$ is positive when $\beta_{0}+\gamma_{0}>0$ ($<0$) if $\alpha-\alpha_{3}>0$ ($<0$) and negative when $\beta_{0}+\gamma_{0}>0$ ($<0$) if $\alpha-\alpha_{3}<0$ (>0).
When a theory tends toward General Relativity, i.e. $\phi\rightarrow \phi_{RG}$, the dynamic of the metric function will be decelerated if:
\begin{equation} \label{34}
\lim_{\phi\rightarrow \phi_{RG}} \omega_{\phi}<2A^{2}\alpha_{0}(\beta_{0}+\gamma_{0})(3+2\omega)^{2}\phi^{-1}
\end{equation}	
Under this condition one can not get inflation at late time. Note that (\ref{34}) has the same form as (\ref{26}) except the introduction of the constant $\beta_{0}+\gamma_{0}$. This comes from the fact that in the $t$ time, all the metric functions appear in each field equations. If we use the three coupling functions of subsection \ref{s62} with equation (\ref{33}), one obtain complex expressions which need numerical investigations to find their zeros.

Since the presence of matter tends to slow down the expansion, one can hypothesize that (\ref{34}) could be a sufficient (but not necessary) condition so that model with matter has a decelerated behaviour in the same circumstances, that is at late time when the theory tends towards a relativistic behaviour.
\section{Conclusions.} \label{s7}
From the form of the coupling function $\omega(\phi)$, we can deduce the qualitative behaviour of the metric functions. It depends on the sign of $d\phi/d\tau$, $d\omega/d\phi$ and the integration constants $\alpha_{0}$, $\beta_{0}$, $\gamma_{0}$. We have studied two things : sign of the first and second derivatives of the metric functions.
 
For the first derivative, the main difficulty is to find the zeros of $\omega_{\phi}$. When $\omega(\phi)$ is a monotonous function of the scalar field, we have eight basic possible behaviours (\{1\}, \{2\}, \{3\}, \{4\}, \{1'\}, \{2'\}, \{3'\}, \{4'\}) for a metric function because $d\phi/d\tau$, $d\omega/d\phi$ and the corresponding integration constants can be positive or negative (2*2*2=8). When $\omega(\phi)$ has one or several extrema, the behaviour of the metric function is a succession of behaviours of types \{i\} + \{i'\}, \{i\} and \{i'\} being the number of  two of the eight basic behaviours, one with $\omega_{\phi}>0$ and the other with $\omega_{\phi}<0$. For the behaviours of type \{1\}, \{1'\}, \{4\} and \{4'\}, a complementary condition has to be fulfilled so that the metric function $a$ ($b$, $c$) has an extremum : the value $2\alpha_{0}$ ($2\beta_{0}$, $2\gamma_{0}$) has to be in the interval in which $d\phi/d\tau$ varies otherwise the metric function is monotone. Or equivalently, a time independent formulation of this condition will be that the value of the scalar field corresponding to $3+2\omega=(2\alpha_{0}A)^{-2}$ ($(2\beta_{0}A)^{-2}$, $(2\gamma_{0}A)^{-2}$) have to belong to the interval in which $\phi$ varies.

For the second derivative of the metric functions in the $\tau$ time, if the condition (\ref{23}) is fulfilled, the dynamic of the metric functions is always accelerated. If it is not the case, we have to examine, for the metric function $a$ for instance, the sign of $\alpha_{1}-\alpha$ and $\alpha_{2}-\alpha$. If these expressions have the same sign, the second derivative of $a$ is positive otherwise it is negative.

In the $t$ time, the dynamic is accelerated if (\ref{32}) is positive and decelerated otherwise. If moreover, the first derivative is positive, we have inflation.
 
With this method we have been able to completely determine, whatever $\tau$, the qualitative form of the metric functions for three different theories. Each of them can be compatible with the value of the PPN parameters at late time if we adjust their parameters. By using the results of subsection \ref{s63} concerning the sign of the second derivative in the cosmic time and numerical calculations, it is also possible to obtain the qualitative form of the metric functions in the $t$ time.

Moreover, if with $\omega\rightarrow +\infty$ and $\omega_{\phi}\omega^{-3}\rightarrow 0$, we want the three metric functions to be increasing and decelerated at late time in the cosmic time, we deduce of the study that we must have: $(\alpha_{0},\beta_{0},\gamma_{0})>0$ and $A$ and $\omega_{\phi}$ must have the same sign, which is positive since $\omega\rightarrow +\infty$ and $\omega_{\phi}<2A^{2}\inf \mbox{[} \alpha_{0}(\beta_{0}+\gamma_{0})$, $\beta_{0}(\alpha_{0}+\gamma_{0})$, $\gamma_{0}(\alpha_{0}+\beta_{0})  \mbox{]}$$(3+2\omega)^{2}\phi^{-1}$ when $\phi$ tends towards $\phi_{RG}$, $\phi_{RG}$ being the smallest value of the scalar field. In these conditions the metric functions have a power law form.

In section \ref{s5}, we have determined the conditions to have 1, 2 or 3 increasing metric functions; in fact, this is a graphic translating of some information contained in the constraint equation of the field equations.

We have studied the simplest anisotropic cosmological model but we hope to extend this method to more complicated ones such as Bianchi types II and V and in more complex situations, i.e. with cosmological constant or potential. The main advantage of such study is to reveal completely the dynamic of the metric functions whatever the form of the coupling function and not only for a particular one or for asymptotic behaviour.

\newpage
\begin{table}
\begin{center}
\begin{tabular}{|c|c|c|c|c|c|}
\hline
Sign of 				& Variation of & Variation of & half-line	& direction of	 & type of \\
($\omega_{\phi}$, $A$, $\alpha_{0}$) & $\phi$ 	& $\alpha(\tau)$ & number 	& the monotone & behaviour \\
				 	& 		& 		& 		&motion of the & number \\
				 	& 		& 		& 		& ($\alpha$, $\Phi$) point & \\
\hline
(+,+,+)	& $\phi^{,}>0$, $\phi^{,,}<0 $	& minimum in $\tau_{1}$		& \{1\}& left to right 	& \{1\} \\
	& 				& when $\Phi=2\alpha_{0}$	& 	& 		& \\
\hline
(+,-,+)	&$\phi^{,}<0$, $\phi^{,,}<0$ 	& increasing			&\{2\}	& left to right 	&\{2\} \\
\hline
(+,+,-)	&$\phi^{,}>0$, $\phi^{,,}<0 $	& decreasing			&\{3\}	& left to right	&\{3\} \\
\hline
(+,-,-)	&$\phi^{,}<0$, $\phi^{,,}<0$	& minimum in $\tau_{1}$	&\{4\}	& left to right 	&\{4\} \\
	& 				& when $\Phi=2\alpha_{0}$	& 	& 		& \\
\hline
(-,+,+)	&$\phi^{,}>0$, $\phi^{,,}>0 $	& maximum in $\tau_{1}$	&\{1\}	& right to left 	&\{1'\} \\
	& 				& when $\Phi=2\alpha_{0}$	& 	& 		& \\
\hline
(-,-,+)	&$\phi^{,}<0$, $\phi^{,,}>0 $	& increasing 			& \{2\} 	& right to left 	&\{2'\} \\
\hline
(-,+,-)	&$\phi^{,}>0$, $\phi^{,,}>0 $	& decreasing 		& \{3\} 	& right to left 	&\{3'\} \\
\hline
(-,-,-)	&$\phi^{,}<0$, $\phi^{,,}>0 $	& maximum in $\tau_{1}$	&\{4\}	& right to left 	&\{4'\} \\
	& 				& when $\Phi=2\alpha_{0}$	& 	& 		& \\
\hline
\end{tabular}
\end{center}
\caption{The eight types of behaviours of  the scalar field and metric function when the coupling constant is a monotone function of the scalar field. Note that the sign of the second derivative of $\phi$ with respect to $\tau$ or $t$ will not be the same. But the signs of all the first derivatives will stay the same.}
\end{table}
\newpage
Figure 1 : solution of the first equation of (\ref{7}) in the $\alpha,\Phi$ plane depending on the sign of $\alpha_{0}$.
\newline
\newline
Figure 2 : the four different physically solutions of the first equation of (\ref{7}).
\newline
\newline
Figure 3 : representation of all the solutions of the equations (\ref{7}) in the $((\alpha,\beta,\gamma),\Phi)$ plane.

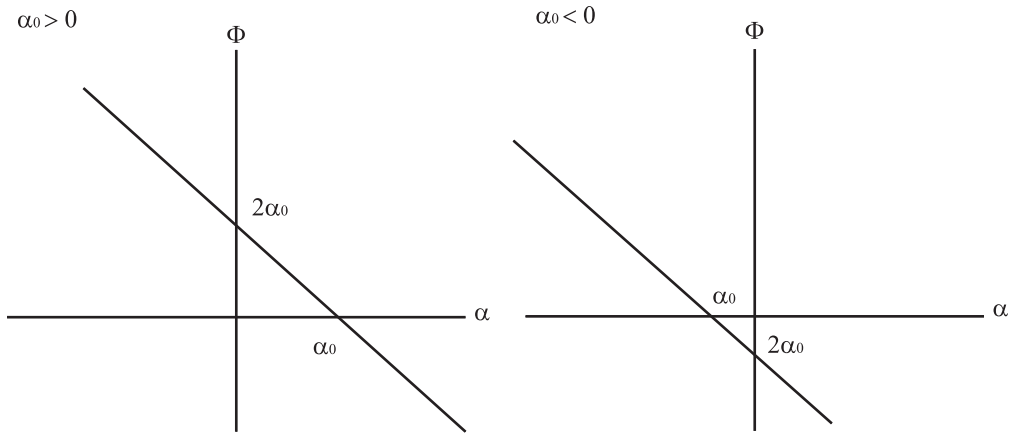
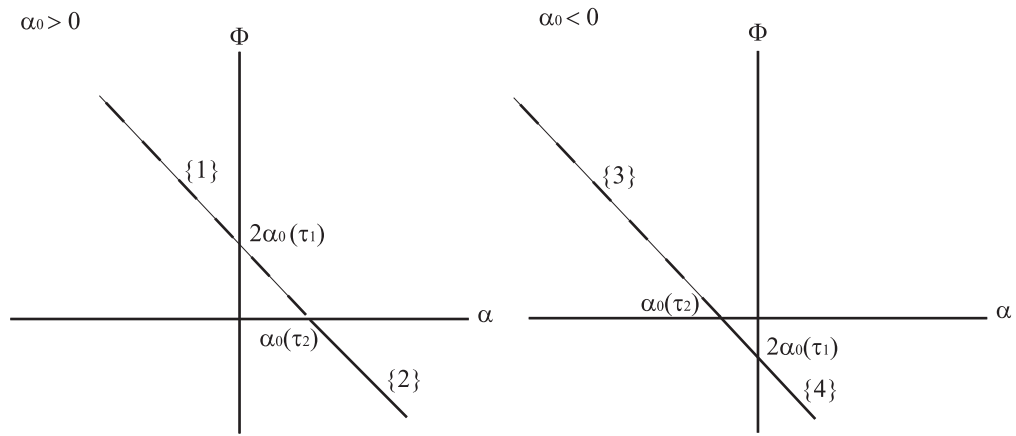
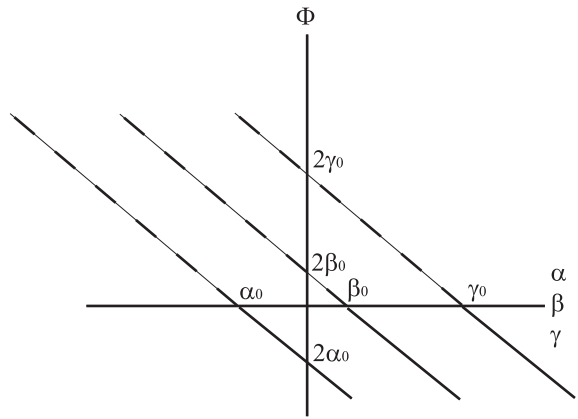
\end{document}